\documentclass[preprint,12pt]{elsarticle}

\usepackage{xcolor}
\usepackage{graphicx}  
\usepackage{dcolumn}
\usepackage{bm}
\usepackage{mathrsfs}
\usepackage{amsmath} 
\usepackage{amssymb}

\newcommand{\icarus}{Icarus}
\newcommand{\apjl}{ApJL}
\newcommand{\apj}{ApJ}
\newcommand{\prd}{PRD}
\newcommand{\prl}{PRL}
\newcommand{\jcap}{JCAP}
\newcommand{\mnras}{MNRAS}
\newcommand{\aap}{A\&A}

\journal{Physics of the Dark Universe}

\begin{document}

\begin{frontmatter}

\title{Limiting the Yukawa Gravity through the Black Hole Shadows of Sgr A* and M87*}

\author[label1,label2]{Yuan Tan}
\author[label1,label2]{Youjun Lu}
\ead{luyj@nao.cas.cn}
\author[label1,label2]{Kunyu Song}
\affiliation[label1]{organization={National Astronomical Observatories, Chinese Academy of Sciences},
            addressline={20A Datun Road},
            city={Beijing},
            postcode={100101},
            state={Beijing},
            country={China}}
\affiliation[label2]{organization={School of Astronomy and Space Sciences, University of Chinese Academy of Sciences},
            addressline={19A Yuquan Road},
            city={Beijing},
            postcode={100049},
            state={Beijing},
            country={China}}

\begin{abstract}
Recently, the \textit{EHT} collaboration unveiled the shadow images of the supermassive black hole (SMBH) M87* and Sgr A*, with angular radii of $42\pm3$\,$\mu$as and $48.7\pm7.0$\,$\mu$as, respectively. These observations are consistent with the shadow of a Kerr black hole in general relativity (GR). Observations of the shadow of SMBHs can be used to test modified gravity theories, including Yukawa gravity, in extremely strong fields. In this paper, we illustrate the shadows of Yukawa black holes, showing that their sizes are significantly influenced by the Yukawa parameters $\lambda$ and $\kappa$. Using the EHT observations of M87* and Sgr A*, we obtain constraints on the Yukawa parameters. For Sgr A*, Keck and VLTI provide different priors on its gravitational radius. The Sgr A* shadow yields $\kappa=-0.04^{+0.09}_{-0.10}$ for $\lambda>1$\,AU with the Keck prior, while $\kappa=-0.08^{+0.09}_{-0.06}$ with the VLTI prior. As $\lambda$ decreases, the constraints weaken, reaching $-0.37<\kappa <0.17$ (Keck prior) and $-0.47<\kappa<0.04$ (VLTI prior) at $\lambda=0.1$\,AU. For M87*, with a mass significantly larger than Sgr A*, this system can only put constraints on $\kappa$ at larger $\lambda$. For $\lambda>1.5\times10^4$\,AU, the \textit{EHT} observation of M87* yields $\kappa=-0.01^{+0.17}_{-0.17}$. No significant deviation from GR is detected in our analysis. Additionally, we explore potential constraints using the next-generation VLBI, like \textit{ngEHT} and the Black Hole Explorer (BHEX), which promise the detection of the second ring of photons. The improved angular resolution and the measurements of the second ring could substantially refine constraints on the Yukawa parameters, enhancing our ability to test deviations from GR in the strong-field regime.
\end{abstract}


\end{frontmatter}

\section{Introduction}
\label{introduction}
 
Over the past decades, General Relativity (GR) has achieved remarkable success. However, certain observational phenomena, such as the accelerated expansion of the universe and the rotation curves of galaxies, challenge GR and often necessitate the introduction of additional components like dark matter and dark energy \citep{2010ARA&A..48..495F, 2005PhR...405..279B, 2014NatPh..10..496S, 2018Galax...6...10D, DeMartino2018PRD, 2020Univ....6..107D, reviewer..suggest1}. To date, no direct evidence for these enigmatic components has been found. Consequently, various alternative theories of gravity have been proposed.

Verlinde proposed that gravity is an entropic force arising from the change of information in a system \citep{2011JHEP...04..029V}, arguing further that dark matter is not a physical entity but rather an apparent effect, i.e., a consequence of baryonic matter \citep{2017ScPP....2...16V}. Recently, this concept has been applied to derive modified Friedmann equations incorporating the effects of a minimal length scale \citep{2023PhLB..83637621J, 2023PhLB..84137916M, 2023PDU....4201270J, 1973ApJ...183..237C}. As shown in \cite{2023PDU....4201318J}, dark matter can be interpreted as a coupling between baryonic matter mediated by a long-range force governed by the Yukawa gravitational potential. This coupling is characterized by the coupling parameter $\kappa$, the wavelength parameter $\lambda$, and the Planck length $l_0$. The modified Friedmann equations are derived using Verlinde’s entropic force framework, based on the holographic scenario and the equipartition law of energy. These equations establish a relationship between the densities of dark matter, dark energy, and baryonic matter. Importantly, dark matter in this framework is not due to unknown particles but is understood as an emergent, apparent effect.

{Yukawa-like terms were first introduced by modifying the Newtonian gravitational potential to explain the flat rotation curves of spiral galaxies without invoking dark matter \citep{Sanders1984}. More recently, such Yukawa-like interactions have been shown to arise naturally in the weak-field limit of various extended theories of gravity. Examples include Scalar-Tensor-Vector gravity \citep{Moffat2006JCAP}, massive gravity theories \citep{Visser1998, Hinterbichler2012}, higher-dimensional theories with Kaluza–Klein compactification \citep{Bars1986, Hoyle2001}, massive Brans–Dicke theories \citep{Perivolaropoulos2010, Alsing2012}, and $f(R)$ gravity \citep{Capozziello2012}, etc.. In addition, similar corrections can also emerge in the context of Verlinde’s emergent gravity framework.}

Recent research has investigated the Yukawa potential across a wide range of scales and environments, from millimeter-scale dual-modulation torsion pendulum experiments \citep{2012PhRvL.108h1101Y} to Solar System dynamics \citep{2007JHEP...10..041I, 2011Icar..211..401K, 2012JHEP...05..073I, 2023arXiv230913106T}, the orbits of S-stars near the Galactic Center \citep{Hees2017PRL, DeMartino2018PRD, 2018PhRvD..97j4068D, 2024PhRvD.109d4047T}, and the dynamics of galaxies \citep{2011MNRAS.414.1301C, 2012ApJ...748...87N, 2024arXiv240401846D} etc. However, past observational limitations have precluded testing the Yukawa potential in regions of extremely strong gravitational fields, such as those just a few Schwarzschild radius away from a supermassive black hole (SMBH). This situation is now changing, as new observational opportunities have emerged, opening the door to exploring the Yukawa potential in these extreme environments.

In 2019, the international \textit{Event Horizon Telescope} (\textit{EHT}) collaboration revealed the first shadow image of SMBH M87*, with a prominent thick ring with an angular diameter of $42\pm 3$\,$\mu$as and a fractional deviation of $\delta = -0.01\pm 0.17$ \citep{2019ApJ...875L...1E, 2019ApJ...875L...2E, 2019ApJ...875L...3E, 2019ApJ...875L...4E, 2019ApJ...875L...5E, 2019ApJ...875L...6E}. Shortly thereafter, \textit{EHT} released the shadow image of Sgr A*, the SMBH at the center of our galaxy, with a central shadow diameter of $48.7\pm 7.0$\,$\mu$as and Schwarzschild shadow deviations of $\delta = -0.04^{+0.09}_{-0.10}$ for the Keck prior and $\delta = -0.08^{+0.09}_{-0.09}$ for the VLTI prior \citep{2022ApJ...930L..12E, 2022ApJ...930L..13E, 2022ApJ...930L..14E, 2022ApJ...930L..15E, 2022ApJ...930L..16E}. These groundbreaking observations of SMBH shadows have ushered in a new era of exploring the properties of black holes and the nature of strong-field gravity at the event horizon scale. 

{Numerous studies have used the current \textit{EHT} observations to constrain a number of gravity theories. Vagnozzi et al. (2023) \citep{Vagnozzi2023} analyzed the EHT Sgr A* image to test a wide range of scenarios, including various regular BHs, string-inspired space-times, violations of the no-hair theorem, and BH mimickers. Psaltis et al. (2020) \citep{Psaltis2020PRL} demonstrated that measurements of the M87* shadow size can impose strong constraints on deviation parameters associated with second post-Newtonian and higher-order corrections to the metric. Other efforts include constraints on black hole charge \citep{Kocherlakota2021PRD, Ghosh2023ApJ}, limits on axion-photon coupling \citep{2022NatAs...6..592C}, tests of Horndeski gravity \citep{Afrin2022ApJ}, bounds on ultralight bosons \citep{Saha2024}, and probes of quantum gravity effects such as loop quantum gravity \citep{Afrin2023ApJ} and quantum-corrected black holes \citep{Vachher2025, Raza2025}. According to these studies, no significant deviations from the predictions of general relativity have been observed. We note here that the Yukawa metric, equally important as other alternative gravity theories, has not been studied in the extreme gravitational environments observed by \textit{EHT}, yet.}

\begin{figure}[htb]
\centering
\includegraphics[width=\textwidth]{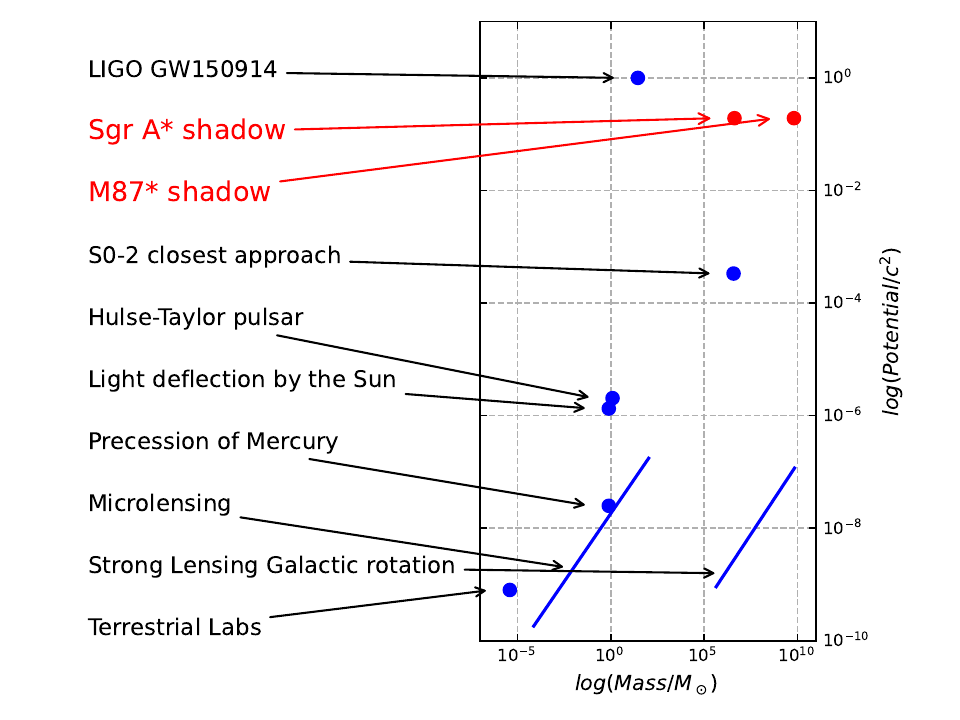}
\caption{
The gravitational potential probed by different tests against the mass of the central body that generates gravity in these tests. The black hole shadow observations explore a new region in this parameter space. The figure is inspired by \cite{2004AIPC..714...29P} and \cite{Hees2017PRL}.
}
\label{mass_potential}
\end{figure}

Furthermore, the ongoing next-generation \textit{EHT} (\textit{ngEHT}) project, being designed to significantly enhance the observational capabilities of the current \textit{EHT} \citep{2019BAAS...51g.256D}. {A space-based extension to the EHT, known as the Black Hole Explorer (BHEX), has recently been proposed \citep{Lupsasca2024}. It will provide a robust detection of the second photon ring (usually called $N=1$ photon in the photon ring community). This structure, dictated solely by unstable photon orbits, depends mostly on the black hole spacetime properties}. Thus, the direct detection of the second photon ring may offer an even more precise constraint on spacetime metrics such as the Yukawa metric. As illustrated in Figure~\ref{mass_potential}, it is evident that observing the shadow allows us to explore a new regime of gravity in extremely strong fields. This system shows significant potential for constraining modified gravity theories. 

This paper is organized as follows. In Section~\ref{shadow}, we introduce the Yukawa metric and calculate the critical curve and the second ring for a Yukawa black hole by direct calculation and the ray-tracing method, respectively. In Section~\ref{constrains}, we use \textit{EHT} observations of M87* and Sgr A* to constrain the Yukawa parameters. Section~\ref{ngEHT constraint} explores the potential constraints on these parameters through the second ring measurements by future telescopes, such as \textit{ngEHT} and BHEX. Finally, we present our main conclusions in Section~\ref{conclusions}. 

\section{Shadows of the Yukawa-like metric} 
\label{shadow}

The general solution of the metric for the case of a static, spherically symmetric source is
\begin{equation}
ds^2=-f(r)dt^2+\frac{dr^2}{f(r)}+r^2(d\theta^2+\sin^2\theta d\phi^2).
\end{equation}
\cite{Gonzalez2023PDU} finds the solution for the Einstein field equations when considering the regular Yukawa-type potential $\Phi(r)=-\frac{Mm}{\sqrt{r^2+\ell_0^2}}\left(1+\kappa e^{-\frac{r}{\lambda}}\right)$ as
\begin{equation}
f(r)=1-\frac{2M}r\left[1+\kappa e^{-\frac r\lambda}(1+\frac{r}{\lambda})\right],
\end{equation}
where $\lambda$ represents the characteristic interaction length of the Yukawa potential and $\kappa$ characterizes the strength of this interaction. In the above equation, we also set $G=c=1$. Since in this paper we only consider systems with scales much smaller than the cosmological scale, the cosmological constant term is neglected. Given that $\ell_0$ is of Planck length order, i.e., $\ell_0 \sim 10^{-35}$\,m \citep{2019PhLB..79734888N}, we have $r \gg \ell_0$, and thus $\ell_0$ can be ignored for the SMBH systems studied in this paper. 

%\TY{We note that such a
The above potential corresponds to the Yukawa-modified force
\begin{equation}
F = - \nabla \Phi(r) \bigg|_{r=R}=-\frac{G M m}{R^{2}}\left[1+\kappa\left(\frac{R+\lambda}{\lambda}\right) e^{-\frac{R}{\lambda}}\right],
\end{equation}
which can be derived by modifying the expression for entropy adopting Verlinde’s entropic force interpretation. According to this framework, when a test particle or excitation moves away from the holographic screen, the magnitude of the entropic force acting on the body is given by $F \Delta x = T \Delta S $ \citep{2011JHEP...04..029V}. Specifically, Newton’s law of gravitation can be recovered from the entropy-area relationship $S = A/4 = \pi R^2$. However, in general, the expression of total entropy can be modified to $S = \pi R^2 + \mathcal{S}(R)$ \citep{2023PDU....4201318J}. If the entropy is set as
\begin{equation}
S=\pi R^{2}-2 \pi \kappa\left(R^{2}+3 \lambda R+3 \lambda^{2}\right) e^{-\frac{R}{\lambda}},
\end{equation}
we can derive the Yukawa-modified force. This demonstrates the self-consistency of this potential within the framework of modified entropy gravity \citep{2023PDU....4201318J, Gonzalez2023PDU}. It is important to note that the entropy correction can be interpreted as a volume-law entanglement contribution to the entropy due to gravitons. From the above equation, we observe that $\kappa$ appears solely in the second term, suggesting that $\kappa$ arises from the entanglement associated with the contribution from volume-law entropy. We now proceed to explore the shadow properties of this modified gravitational law.

To analyze the shadow properties of the Yukawa-like black hole, we employ two kinds of methods as detailed below. The first one is to calculate the critical curve of the black hole directly, and the second one is to calculate the second ring through ray-tracing techniques.

\subsection{Critical curve}
With decreasing impact parameters of photons, they complete an increasing number of orbits around the black hole in its vicinity before escaping to a distant observer. The trajectories of these photons on the observation screen converge to a well-defined boundary, known as the \textit{critical curve}. This critical curve is independent of the accretion model and is solely determined by the metric. In this Section, we will calculate the critical curve from the Yukawa metric directly.

Without loss of generality, we consider photon trajectories confined to the equatorial plane, setting $\theta=\pi/2$. For a spherically symmetric static metric, the Killing vectors of spacetime provide the constants of motion: energy $E=-p_t$ and angular momentum $L=p_\phi$. For photons, the radial motion equation can be derived from the condition $g_{\mu\nu}p^\mu p^\nu=0$ as 
\begin{equation}
\frac{1}{E^2}\left (\frac{\mathrm{d}r}{\mathrm{d}x}\right)^2=\mathcal{R}(r)=-\frac{(\frac{1}{g_{tt}} + \frac{\mathscr{L}^2}{r^2})}{g_{rr}},
\end{equation} 
where $\mathscr{L}=L/E$ is the angular momentum rescaled by energy.

The critical curve is determined by the innermost bound circular geodesic, which satisfies the following criteria, i.e., 
\begin{equation}
\mathcal{R}(r)=0, \ \ \ \mathcal{R}'(r)=0,
\label{eq:RRp}
\end{equation}
where $\mathcal{R}'(r) \equiv d\mathcal{R}(r)/d r$ is the first derivative of $\mathcal{R}$ with respect to the radius. These conditions allow us to determine the radius $r_{\rm ph}$ of the innermost bound circular orbit of photons and the corresponding energy-rescaled angular momentum $\mathscr{L}_{\rm ph}$. From the first condition, we have
\begin{equation}
\begin{split}
|\mathscr{L}_{\rm ph}|=\sqrt{-\frac{r_{\rm ph}^2}{g_{tt}(r_{\rm ph})}}=\sqrt{\frac{r_{\rm ph}^2}{1-\frac{2M}{r_{\rm ph}}\left[1+\kappa e^{-\frac{r_{\rm ph}}{\lambda}}\left(1+\frac{r_{\rm ph}}{\lambda}\right)\right]}}.    \\
\end{split}
\end{equation}
However, the second condition, i.e., $\mathcal{R}'(r)=0$, is analytically intractable due to its complexity. Alternatively, we employ the Runge-Kutta method (RK45) \citep{RK45_1,RK45_2} to numerically solve $r_{\rm ph}$. Then the shadow radius on the distant observer screen is determined by \citep{1973ApJ...183..237C}
\begin{equation}
r_{\rm sh}=-\mathscr{L}_{\rm ph}.
\end{equation}

Note here that Equation~\eqref{eq:RRp} may not have a solution when $\kappa \ll 0$, with which $g_{tt}(r_{\rm ph})>0$. In such cases, no stable circular orbit exists around the Yukawa black hole, and consequently, no shadow image can form. These scenarios can be ruled out by the observed \textit{EHT} black hole ring images.

\subsection{Second photon ring} 
\label{ray-tracing}

The second ring in a black hole shadow is also of great importance for constraining the black hole metric \citep[][]{2022ApJ...927....6B, 2019PhRvD.100b4018G,2020SciA....6.1310J}. However, the method described above can only determine the radius of the critical curve, and the ray-tracing technique is required for calculating the radius of the second photon ring. In the following, we introduce the ray tracing method, which was first developed for the Kerr metric \citep[e.g.,][]{1975ApJ...202..788C, 1992MNRAS.259..569K, 1994ApJ...421...46R, 2015ApJ...809..127Z, 2016ApJ...827..114Y} and later for various metrics from modified gravities or phenomenological arguments \citep[e.g.,][]{2023PhRvD.108l4054D, 2024EPJC...84..441S, 2024JCAP...01..059J, 2023PhRvD.107l4026S}. 

\subsubsection{Motion equations of photons}

Consider a photon propagating along the null geodesic from the vicinity of an SMBH to a distant observer. Since most photons cannot reach the observer, tracing photons starting from the black hole can be computationally inefficient. Instead, photons can be traced backward from the observer to the vicinity of the black hole (with the Yukawa metric) and then integrate the radiative transfer equation forward, from the black hole vicinity to the telescope.

Since the metric is spherically symmetric, we set $\theta=\pi/2$ to reduce the problem to two spatial dimensions. The motion equations of a photon can be derived from the Lagrangian
\begin{equation}
\mathcal{L}=\frac{1}{2}\big(-f(r)\dot{t}^2+\frac{1}{f(r)}\dot{r}^2+r^2\dot{\phi}^2\big).
\end{equation}
This Lagrangian does not depend explicitly on the coordinates $t$ and $\phi$, yielding conserved momenta as
\begin{eqnarray}
p_t & = & \frac{\partial{\mathcal{L}}}{\partial{\dot{t}}}=-E,\\
p_\phi & = & \frac{\partial{\mathcal{L}}}{\partial{\dot{\phi}}}=L,
\end{eqnarray}
with $E$ and $L$ representing the energy of the photon at infinity and the angular momentum of the photon in the $\phi$ direction. The corresponding geodesic equations of motion of $t$ and $\phi$ directions are then written as
\begin{eqnarray}
\dot{t}& = & f(r)^{-1}E,  \\
\dot{\phi}& = & L/r^2.
\end{eqnarray}
For the radial direction, the canonical momentum is
\begin{equation}
p_r=\frac{\partial{\mathcal{L}}}{\partial{\dot{r}}}=\frac{\dot{r}}{f(r)}.
\end{equation}

Using the Hamiltonian $\mathcal{H} = x^\mu p_\mu - \mathcal{L}$, the equation of motion for the covariant radial component of the four-momentum can be expressed as
\begin{equation}
\begin{split}
    &\dot{p}_r=\frac{\partial \mathcal{H}}{\partial r}=\frac{1}{\lambda \left( -2 \kappa e^{-\frac{2}{\lambda}} + (-2 + r) \right)^2 r^3 } \\
    &\quad\bigg[-4 \kappa^3 p_{r}^2 r (\lambda + r) e^{-\frac{3r}{\lambda}} + 4 \kappa^2 \bigg( p_{r}^2 (-2 + r) r^2  \\
    &\quad+ \lambda \left(p_{r}^2 (-3 + r) r \right) \bigg) e^{-\frac{2r}{\lambda}} + \lambda \left(  - r \left( p_{r}^2 (-2 + r)^2 + E^2 r^2 \right) \right)  \\
    &\quad- \kappa \bigg( r^2 \left( p_{r}^2 (-2 + r)^2 + E^2 r^2 \right) + \\
    &\quad\lambda \left( E^2 r^3 + p_{r}^2 r (12 - 8r + r^2) \right) \bigg) e^{-\frac{r}{\lambda}}  \\
    &\quad+ \lambda L^2 \left( -2 \kappa + e^{\frac{r}{\lambda}} (-2 + r) \right)^2 e^{-\frac{2r}{\lambda}} \bigg].
\end{split}
\end{equation}
The complete set of geodesic equations of motion for the four variables $(r, \phi, t, p_r)$ is then determined by these expressions.

The radius of the second photon ring is also an important feature of the black hole shadow, which is described in detail in this section. The impact parameter in the observer’s image frame, $b$, can be calculated as \citep{1983mtbh.book.....C}
\begin{equation}
b=\frac{L}{\sin\theta_{\mathrm{obs}}}=L.
\end{equation}
We set the photon energy to unity and the observer at $r_{\rm obs}=10^6 M$, which is sufficiently large. Thus the initial conditions for a photon with impact parameter $b$ in the observer's sky are
\begin{equation}
(t,r,\theta,\phi)=\left(0,r_{obs},\pi/2,\arccos\left(\frac{b}{r_{obs}}\right)\right),
\end{equation}
and
\begin{equation}
(p_t,p_r,p_\theta,p_\phi)=\left(-1,\sqrt{\frac{-g^{tt}-b^2 g^{\phi\phi}}{g^{rr}}},0,b\right).
\end{equation}
The radial momentum $p_r$ is derived from the normalization of the four-momentum.

We numerically integrate these equations backward from the observer to the vicinity of the black hole. The integration terminates when the photon either reaches the event horizon or escapes to a sufficiently large radius where the emissivity becomes negligible. This approach ensures efficient and accurate computation of photon trajectories and the resulting shadow properties.

\subsubsection{Emission model for accretion flows}

\begin{figure}
\centering
\includegraphics[width=\textwidth]{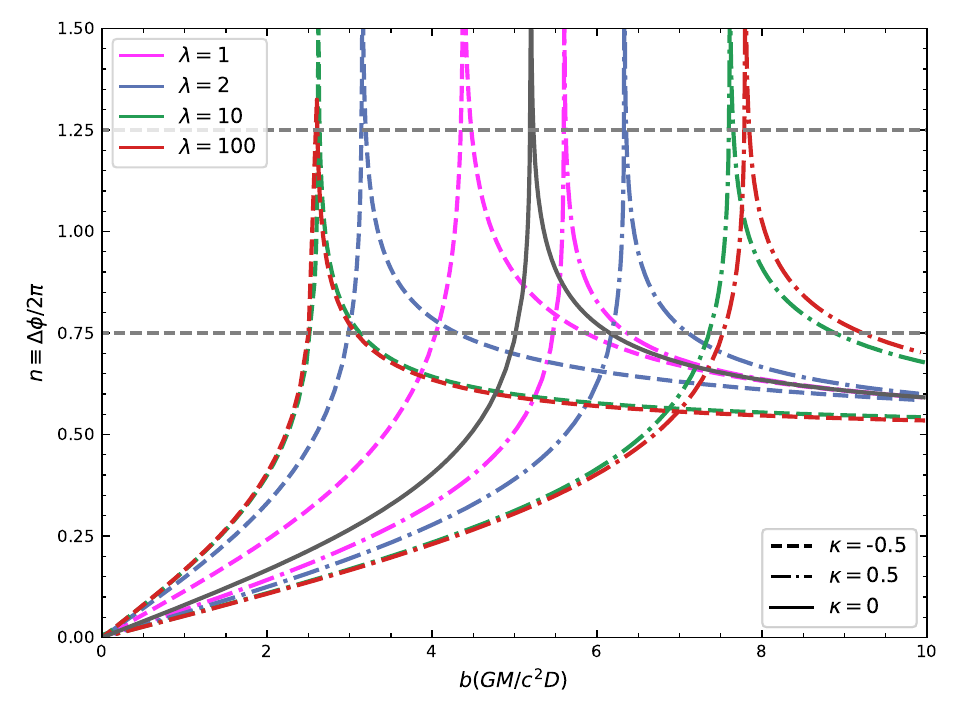}
\caption{
The total number of orbits, $n\equiv\phi/2\pi$, as a function of the impact parameter $b$ for varying Yukawa parameters $\lambda$ and $\kappa$. The gray dashed line represents the critical value $n=3/4$ and $n=5/4$. Those photons with the orbits numbers $3/4<n<5/4$ constitute the second photon ring. 
}
\label{OrbitNum}
\end{figure}

\begin{figure*}[!htp]
\centering
\includegraphics[width=0.89\textwidth]{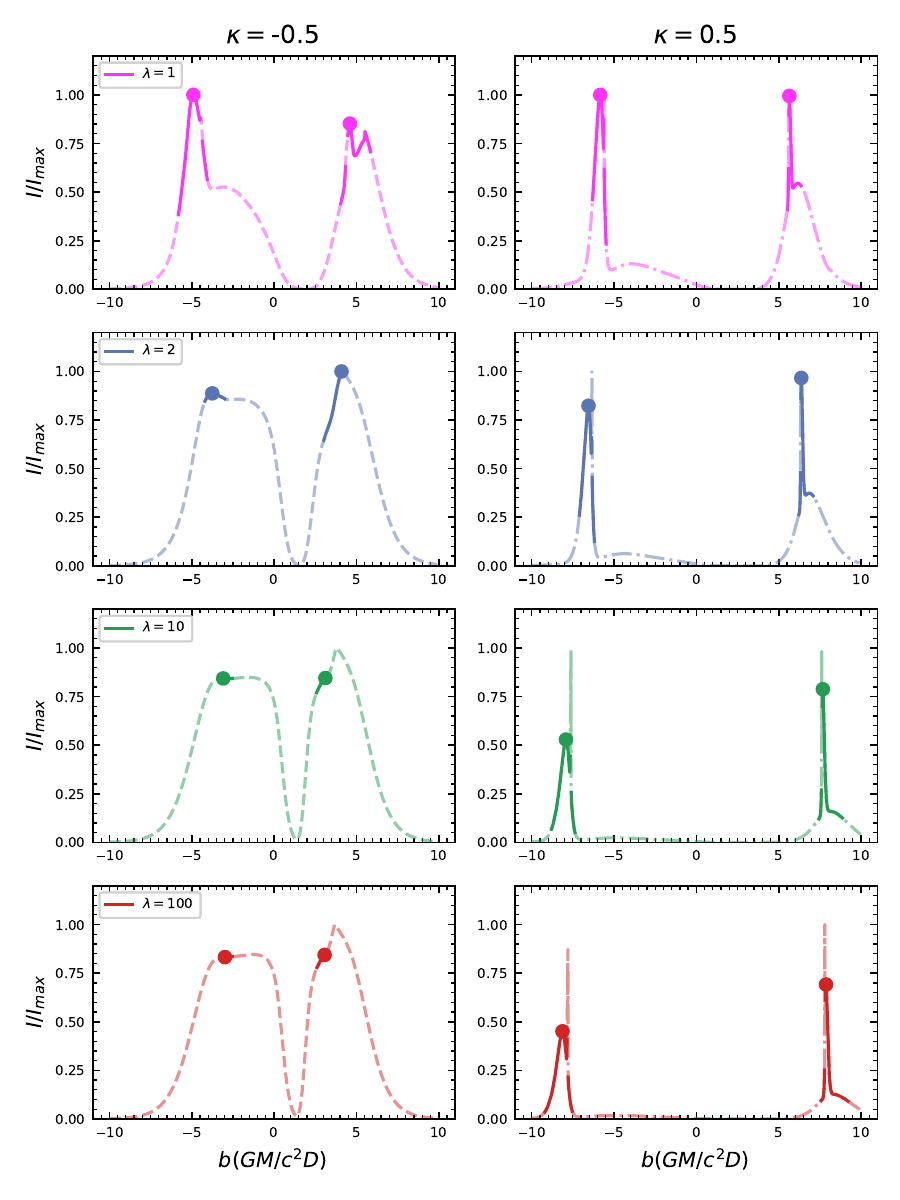}
\caption{
The intensity profiles of the radio emission resulting from the Yukawa metric with different parameters $\lambda$ and $\kappa$. The solid part of each curve represents the intensity from the second photon ring which is composed of photons who satisfy $3/4<n<5/4$, while the filled dots represent the peaks of the second photon rings. Left and right columns show the cases with $\kappa=-0.5$ and $0.5$, respectively. Top to bottom panels show the cases with $\lambda=1$, $2$, $10$, and $100$, respectively.
}
\label{intensity_profile}
\end{figure*}

The emission of Sgr A* at $230$\,GHz is dominated by thermal synchrotron radiation \citep{2000ApJ...541..234O}. Similarly, thermal synchrotron radiation can also account for the observed emission at this frequency in M87* \citep[e.g.,][]{2016A&A...586A..38M, 2016MNRAS.457.3801P, 2018ApJ...864..126R,2019MNRAS.486.2873C, 2019ApJ...875L...5E}. So we consider only thermal synchrotron emission in our analysis. Based on current understanding, the magnetically arrested disk (MAD) model can produce a black hole shadow that is most consistent with \textit{EHT} observations \citep{2022ApJ...930L..16E}. Therefore, we adopt the MAD model as described by \cite{2022NatAs...6..592C} to simulate the disk configuration of Sgr A* and M87*. Note that this emission model is formulated assuming a GR background. We assume that accretion flows in the Yukawa metric behave similarly to those in GR, given that the Yukawa strength parameter $\kappa$ has already been restricted to a narrow range, even on the scale of the shadow systems considered in this paper \citep{2012PhRvL.108h1101Y, 2007JHEP...10..041I, 2011Icar..211..401K, 2012JHEP...05..073I, 2023arXiv230913106T,Hees2017PRL, DeMartino2018PRD, 2018PhRvD..97j4068D, 2024PhRvD.109d4047T,2011MNRAS.414.1301C, 2012ApJ...748...87N, 2024arXiv240401846D}. This assumption allows us to apply the existing GR-based emission model with reasonable confidence. A full Yukawa-MHD simulation would give a better description of the emission model, but that would be computationally expensive and beyond the scope of this paper.

For simplicity, we assume that the plasma within the innermost stable circular orbit (ISCO), $r_\text{ISCO}$, is in free fall, while the disk outside ($r_\text{ISCO}$) rotates with sub-Keplerian velocities. The ISCO radius is numerically determined from the equation of motion. The outer edge of the disk is set at $r_\text{out} = 50r_g$, where $r_{\rm g} = GM/c^2$.

The angular frequency due to free fall is given by $\Omega_\text{FF} = -g_{t\phi}/g_{\phi\phi}$, while the Keplerian angular frequency is $\Omega_\text{K}=\frac{-g_{t\phi,r}-\sqrt{(g_{t\phi,r})^2-g_{tt,r}g_{\phi\phi,r}}}{g_{\phi\phi,r}}$. We define the sub-Keplerian motion of our accretion disk model as
\begin{equation}
\Omega=\Omega_\text{K}+(1-K)(\Omega_\text{FF}-\Omega_\text{K}),
\end{equation}
where the Keplerian factor $K$ is set as $0.5$. The free fall radial velocity is $u^r_\text{FF}=-\sqrt{|-(1 + g^{tt})g^{rr}|}$. For the setup of the sub-Keplerian rotation of the outer disk, the radial velocity $u^r$ satisfies $u^r=(1 - K) u^r_\text{FF}$. The remaining components of the four-velocity are determined as $u^t=\sqrt{(1+(u^r)^2 g_{rr})/(-g_{tt}-\Omega^2 g_{pp})}$ and $u^\phi = u^t \Omega$. 

The spatial distributions of the electron temperature $T_\text{e}$ and thermal electron density $n_\text{e}$ are modeled by a hybrid profile, that is, 
\begin{equation}
\begin{split}
&n_\text{e}=n^\text{0}_\text{e,th} r^{-\alpha}\exp^{-\beta}, \\
 &T_\text{e}=T^0_\text{e} r^{-\gamma},
\end{split}
\end{equation}
where $\alpha = 1.1$ and $\gamma = 0.84$ are adopted to describe the radial dependence, according to the vertically averaged density and temperature profiles found in \cite{2003ApJ...598..301Y}, and 
\begin{equation}
\beta = \frac{1}{2H^2}\frac{z^2}{x^2} \ \ \ (\text{where } z\equiv r \cos \theta, x \equiv r \sin \theta),
\end{equation}
where $\theta$ is the angle measured relative to the normal direction of the accretion disk. The geometry of the accretion flow is then controlled by the value of $H$ (the thickness of the disk), which is set to $0.3$. This setup is consistent with the MAD-like configuration in \cite{2022NatAs...6..592C}. The normalization of the electron density $n_{e,th}^0$ and temperature $T_\text{e}^0$ is set to be $3 \times 10^7$\,cm$^{-3}$ and $3 \times 10^{11}$\,K, ensuring that the total flux density at $230$\,GHz is about $1-3$\,Jy and the magnetic field strength is consistent with \textit{EHT} estimates \citep{2022ApJ...930L..12E, 2022ApJ...930L..14E, 2022ApJ...930L..15E, 2019ApJ...875L...1E, 2019ApJ...875L...4E, 2019ApJ...875L...6E}. 

It is important to note that the coordinate system used here differs from the one in the previous subsection. A rotational transformation determined by the inclination angle $\theta_{\rm disk}$ must be applied for our calculations. For M87*, the inclination angle of the disk is set to $\theta_{\rm disk}=17^\circ$ \citep{2019ApJ...875L...5E}, consistent with \textit{EHT} observations. For Sgr A*, we adopt $\theta_\text{disk} = 45^\circ$, as the \textit{EHT} observations suggest a low inclination angle but do not exclude alternative configurations \citep{2022ApJ...930L..12E,2022ApJ...930L..16E}. Note that variations in this parameter do not significantly affect the results.

For unpolarized emission, we use angle-averaged thermal synchrotron emissivity, accurate to within $2.6$\% for relevant temperatures and frequencies \citep{1996ApJ...465..327M}, i.e.,
\begin{equation}
j_\nu=\frac{n_\text{e}e^2}{\sqrt{3}cK_2(1/\theta_\text{e})}\nu M(x_M),
\end{equation}
where $x_M\equiv\frac{2\nu}{3\nu_b\theta_\text{e}^2}$, $\nu=\frac{\nu_\text{obs}}{\gamma}=\nu_\text{obs}\frac{p_\mu u^\mu|_0}{p_\mu u^\mu|_\infty}$, $p_\mu$ and $u^\mu$ are the covariant four-momentum and the four-velocity of a fluid particle, and $M(x_M)$ is given by
\begin{equation}
\begin{split}
       &M(x_M)= \\
       &\frac{4.0505~a}{x_M^{1/6}}\left(1+\frac{0.40~b}{x_M^{1/4}}+\frac{0.5316~d}{x_M^{1/2}}\right)\times \exp(-1.8899 x_M^{1/3}) 
\end{split}
\end{equation}
with the best-fit coefficients $a$, $b$, and $d$ for different temperatures are given in \cite{1996ApJ...465..327M}. The cyclotron frequency is $\nu_b\equiv eB/2\pi m_\text{e}c$, the dimensionless electron temperature is $\theta_\text{e}\equiv kT_\text{e}/m_\text{e}c^2$, and $K_2(x)$ is the modified Bessel function of the second kind.

The magnetic field strength $B$ is assumed to be in approximate equipartition with the ions,
\begin{equation}
\frac{B^2} {8\pi} = \epsilon n_\text{e} \frac{m_\text{p}c^2r_g}{6r} , 
\end{equation}
where $m_\text{p}$ is the mass of proton and $\epsilon = 0.1$, as adopted in \cite{2011ApJ...738...38B,2011ApJ...735..110B,2016ApJ...820..137B,2018ApJ...863..148P}.

The self-absorbed synchrotron emission in this part of the spectrum is dominated by $S=\frac{j_\nu}{\alpha_\nu}$ \citep{2000ApJ...541..234O}. The synchrotron absorption coefficient $\alpha_\nu$ is related to the emissivity via Kirchhoff's law,
\begin{equation}
\alpha_{\nu}=j_{\nu}/B_\nu(T),
\end{equation}
where $B_\nu(T)=\frac{2h\nu^3}{c^2}\frac1{e^{h\nu/kT}-1}$ is the blackbody source function.

\subsubsection{Second photon ring}

Although the direct calculation method struggles to identify higher-order rings, the ray-tracing method provides a straightforward and efficient approach for illustrating rings of any order by tracking the orbital number for each ray. Figure~\ref{OrbitNum} shows the total number of orbits, $n\equiv\phi/2\pi$, as a function of the impact parameter $b$. The total number of orbits describes how many circles a photon rotates around the black hole before it escapes to a distant observer. The singularity in the plot indicates the position of the critical curve. The second photon ring is determined by those photons that intersect the plane of the disk twice outside the horizon. Note that ``straight line motion” would correspond to $\Delta\phi/2\pi = 1/2$, we select those ring photons by requiring \citep{2019PhRvD.100b4018G}: 
\begin{equation}
3/4 < \Delta\phi/2\pi < 5/4.
\end{equation}

Figure~\ref{OrbitNum} clearly demonstrates that the Yukawa parameters $\lambda$ and $\kappa$ influence the positions of both the second photon ring and the critical curve. The radius of the second photon ring is then determined by identifying the two peaks of the intensity profile, as shown in Figure~\ref{intensity_profile}.

\section{Constraints from \textit{EHT} observations}
\label{constrains}

The \textit{EHT} images of both M87* and Sgr A* reveal a central dark area surrounded by a bright ring-shaped structure. This ring is composed of a series of asymptotically bright photon rings spiraling towards the dark area, predominantly determined by the background spacetime metric. The black hole shadow silhouette, representing the boundary, is expected to be independent of various astrophysical effects. 

The fact that black hole shadows encode information of the spacetime in extremely strong gravitational field suggests that they can be used to test alternative metrics, including the Yukawa metric.

As demonstrated in the previous section, the Yukawa parameters $\lambda$ and $\kappa$ have a significant impact on the size of the black hole shadow and make the shadow deviate from the Schwarzschild shadow. Therefore, the \textit{EHT} observations M87* and Sgr A* can provide unique constraints on the Yukawa parameters $\lambda$ and $\kappa$.

\subsection{Constraints from Sgr A*}

To quantify the difference between the observed shadow of Sgr A* and the theoretical Schwarzschild shadow, the deviation parameter has been obtained as \citep{2022ApJ...930L..12E,2022ApJ...930L..17E}:
\begin{equation}
\delta \equiv  \frac{d_{\rm sh}}{d_{\rm sh,Sch}}-1 = \frac{d_{\rm sh}}{6\sqrt{3}\theta_{\rm dyn}}-1, 
\end{equation}
where $d_{\rm sh}$ is the observed angular diameter of the shadow, $\theta_{dyn} \equiv GM/Dc^2$ represents the gravitational radius inferred from the stellar dynamics. The values of $d_{\rm sh}$ are determined solely from \textit{EHT} observation. It is important to note that the diameter of the ring with the highest intensity in the image is slightly larger than the diameter of the black hole shadow $d_{\rm sh}$. This discrepancy arises from factors such as plasma properties, spacetime and measurement biases, etc. The \textit{EHT} collaboration accounted for these effects by performing extensive simulations across various spacetime geometries and plasma models to estimate formal measurement uncertainties and calibration factors. Their analysis yielded $d_{\rm Sh} = 48.7 \pm 7.0$ \,$\mu$as \citep{2022ApJ...930L..17E}. 

\begin{figure*}
\centering
\includegraphics[width=\textwidth]{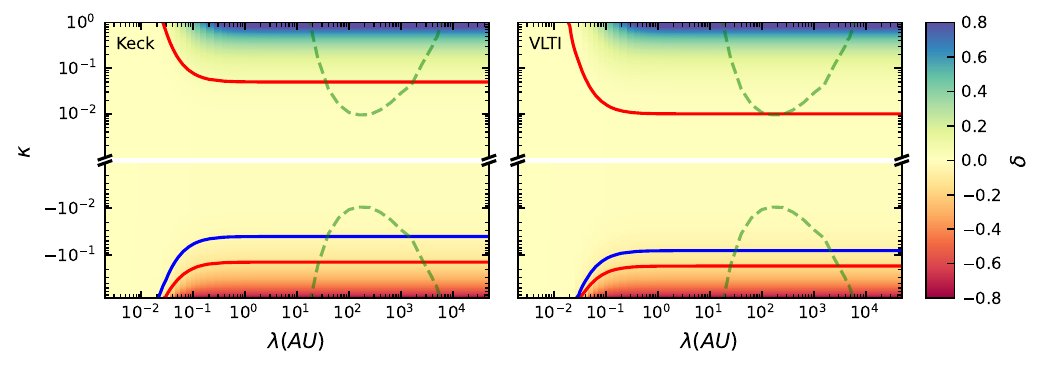}
\caption{
Constraints on $\kappa$ for varying $\lambda$ from \textit{EHT} Sgr A* results of Schwarzschild shadow deviation $\delta$. Based on Keck and VLTI priors, the \textit{EHT} gives two different measurements for $\delta$: $\delta = -0.04^{+0.09}_{-0.10}$ for Keck prior and $\delta = -0.08 ^{+0.09}_{-0.09}$ for VLTI prior. The blue lines correspond to $\delta = -0.04$ for the Keck prior and $\delta=-0.08$ for the VLTI prior, while the red lines show the upper and lower bounds for $\kappa$. For comparison, the constraints obtained from the motion of S-stars are shown by the green dashed lines \citep{2024PhRvD.109d4047T}.
}
\label{constrain_SgrA*}
\end{figure*}

To estimate $\theta_{dyn}$, it is essential to determine both the mass and distance of Sgr A*. The Keck and VLTI teams provided these measurements by analyzing the dynamics of the central stellar cluster within the innermost $10''$ of the Galactic center independently. In the latest publication on the measurement of the gravitational redshift by Keck \citep{2019Sci...365..664D}, they reported a distance of $R_0 = 7959 \pm 59 \pm 32$\,pc and a black hole mass of $M = (3.975 \pm 0.058 \pm 0.026) \times 10^6$\,$M_\odot$. Meanwhile, the VLTI team summarized their recent findings in \cite{2021A&A...647A..59G}, reporting $R_0 = 8277 \pm 9 \pm 33$\,pc and $M = (4.297 \pm 0.012 \pm 0.04) \times 10^6$\,$M_\odot$. Based on these two individual observation, the \textit{EHT} derived two separate constraints for $\delta$: $\delta = -0.04^{+0.09}_{-0.10}$ with the Keck prior and $\delta = -0.08 ^{+0.09}_{-0.09}$ with the VLTI prior.

Because the value of $\delta$ has already accounted for uncertainties when measuring the mass and distance of the black hole, we adopt the Sgr A* mass estimation $M=3.975 \times 10^6$ \,$M_\odot$ for the Keck prior and $M=4.297 \times 10^6$ \,$M_\odot$ for the VLTI prior, respectively. Using these values, the theoretical Yukawa deviation, $\delta_{Yukawa}$, is calculated as:
\begin{equation}
\delta_{\rm Yukawa}=\frac{r_{\rm sh}}{3\sqrt{3}\frac{GM}{c^2}}-1
\end{equation}

Figure~\ref{constrain_SgrA*} illustrates the $\delta_{Yukawa}$ for the Yukawa metric with varying parameters $\lambda$ and $\kappa$. It is evident that $\lambda$ and $\kappa$ significantly influence the value of $\delta_{Yukawa}$. With fixed $\lambda$, reducing $\kappa$ leads to a smaller $\delta_{Yukawa}$, while increasing $\kappa$ enlarges it. Furthermore, the influence of $\kappa$ becomes more pronounced when $\lambda$ increases. When $\lambda$ reaches values on the order of hundreds of Schwarzschild radii, this effect persists at a consistent level. These features can help us to constrain the Yukawa parameters with the \textit{EHT} observations.

The resulting constraints are illustrated in Figure~\ref{constrain_SgrA*}. For $\lambda>1$\,AU, the Keck prior Sgr A* shadow measurement yields $\kappa=-0.04^{+0.09}_{-0.10}$, while the VLTI prior yields $\kappa=-0.08^{+0.09}_{-0.06}$. These constraints surpass those of S-star measurements by more than an order of magnitude (green dashed line) \citep{2024PhRvD.109d4047T}. As $\lambda$ decreases, the constraints on $\kappa$ become weak, since the Schwarzschild radius of Sgr A* is $\sim 0.08$\,AU. At $\lambda=0.1$\,AU, we can find $-0.37<\kappa<0.17$ for the Keck prior and $-0.47<\kappa<0.04$ for the VLTI prior. 

Tan \& Lu (2024) \cite{2024PhRvD.109d4047T} argued that this increase occurs when $\lambda$ is comparable in magnitude to the apocenter distance and is a result of the counteraction between two opposing precessions near the pericenter and apocenter stages. However, the situation is fundamentally different for the black hole shadow system. In this case, all photons originate near the black hole and eventually reach the observer. As a result, this system probes the effects of the entire spacetime, rather than being limited to a localized region, as is the case with S-star observations.

\subsection{Constraints from M87*}

\begin{figure}
\centering
\includegraphics[width=\textwidth]{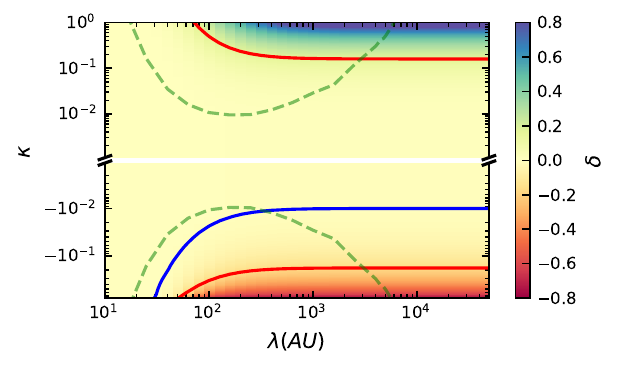}
\caption{
Legends are similar to that for Figure~\ref{constrain_SgrA*}, except for M87* with $\delta=-0.01\pm 0.17$.
}
\label{constrain_M87*}
\end{figure}

For M87*, the \textit{EHT} has introduced a similar fractional deviation as:
\begin{equation}
\delta \equiv  \frac{\theta_g}{\theta_{\rm dyn}}-1
\end{equation}
where $\theta_g$ represents the angular gravitational radius derived from the GRMHD simulations and the \textit{EHT} observation, while $\theta_{\rm dyn}$ is inferred from stellar dynamics \citep{2011ApJ...729..119G}. The \textit{EHT} collaboration reported $\delta=-0.01\pm 0.17$ \citep{2019ApJ...875L...6E}. 

We adopt stellar kinematics inferred M87* mass $M=6.6\times10^9$ \,$M_\odot$ \citep{2011ApJ...729..119G} as a prior to constrain the Yukawa parameters and show the results in Figure~\ref{constrain_M87*}. As the mass of M87* is significantly greater than that of Sgr A*, this system is sensitive to constraining $\kappa$ at large values of $\lambda$. For $\lambda>1.5\times10^4$\,AU, the \textit{EHT} observation of M87* yields $\kappa=-0.01^{+0.17}_{-0.17}$.

\subsection{The influence of spin}

\begin{figure}
    \centering
    \includegraphics[width=\textwidth]{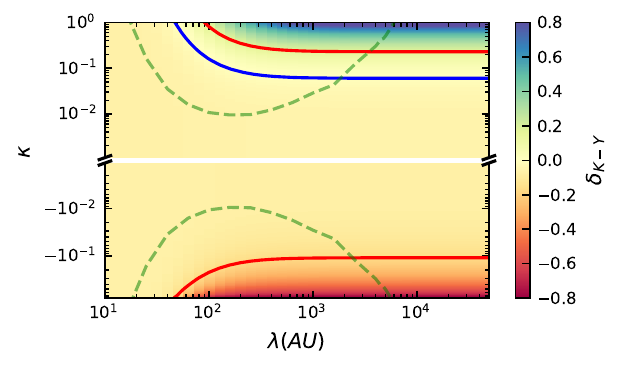}
    \caption{Legends are similar to that for Figure~\ref{constrain_SgrA*}, except for spin $a=0.99$ and inclination $\theta=17^\circ$, and $\delta_{K-Y}$ defined by Equation~\ref{delta_KY} is $ \delta_{K-Y} = -0.01\pm 0.17$.}
    \label{fig:M87*_includeSpin_0.99}
\end{figure}

{We restrict ourselves to the static spherically symmetric metric in this paper, i.e., neglecting the effect of spin, since the rotating version of the Yukawa metric is still not available now. We note, however, there is currently no observational evidence to suggest that either Sgr A* or M87* is a nonrotating object or simply a Schwarzschild black hole. 

For Sgr A*, there is a disagreement on the value of its spin if assuming it is a Kerr black hole. Observations of the alignment of the orbital planes of the S-stars today require Sgr A*'s spin to be very low, $a \lesssim 0.1$ \citep{Fragione2020ApJ,Fragione2022ApJ}, but the EHT images are in principle consistent with a large spin \citep{EHT2023I,EHT2023II}. The constrains on the inclination are even more uncertain \citep{EHT2023I,EHT2023II}. As for M87*, there is indeed a tendency to high spin in recent measurements if assuming it is a Kerr black hole. Based on EHT data, \cite{Tamburini2020MN} and \cite{Drew2025ApJ} estimated the spin parameter as $a=0.90\pm0.05$ and $a\sim0.8$, respectively. The inclination $i$ is approximately $17^\circ$ \citep{2019ApJ...875L...6E, Tamburini2020MN}.  

Although the rotating version of the Yukawa metric is not available now, we give a simple estimate about the influence of the spin on constraining the Yukawa metric. It is well known that the size of the shadow of a Kerr black hole depends only weakly on spin. For all black-hole spins and viewing inclination angles, the average radius of the shadow can be approximated by \citep{Chan2013ApJ}
\begin{equation}
\langle R \rangle \simeq R_0 + R_1 \cos(2.14i - 22.9^\circ),
\end{equation}
where $i$ is the viewing angle, defined as the angle between the line of sight and the spin axis of the black hole, and the two coefficients $R_0$ and $R_1$ are given by
\begin{equation}
\begin{split}
    &R_0 = (5.2 - 0.209a + 0.445a^2 - 0.567a^3), \\
    &R_1 = \left[0.24 - \frac{3.3}{(a - 0.9017)^2 + 0.059}\right]\times 10^{-3}.
\end{split}
\end{equation}
The deviations for the average shadow radius of a spinning black hole from that of a nonspinning one is only
%in Kerr black holes are always small (
$\lesssim6\%$.
%).
Thus, we assume that the deviation for the shadow diameter of a Kerr-Yukawa black hole from a spherical Schwarzschild black hole could be approximated as
\begin{equation}
\delta_{K-Y} \simeq \left(\frac{r_{\rm sh}}{3\sqrt{3}}-1\right)+\left(\frac{\langle R \rangle}{3\sqrt{3}}-1 \right).
\label{delta_KY}
\end{equation}
According to the above equation, we estimate the deviations by simply assuming $a=0.99$ and $i=17^\circ$ to show the influence of the spin, where $a=0.99$ may introduce the largest effect of spin on the shadow size, and $i=17^\circ$ is adopted according to the inclination measurement of M87*. The results are shown in Figure~\ref{fig:M87*_includeSpin_0.99}, which shows that the result does not change significantly even when spin is very high. For $\lambda>1.5\times10^4$\,AU, the \textit{EHT} observation of M87* yields $\kappa=0.05^{+0.17}_{-0.17}$. And we have tested that the value of the inclination $i$ do not affect the result significantly. This result could be further refined in the future work by considering the rotating Yukawa metric, which is perhaps not easy to directly derived from first principle and is beyond the scope of this work.
}

\section{Future constraints from next generation VLBI observations}
\label{ngEHT constraint}

\begin{figure*}[!htp]
\centering
\includegraphics[width=\textwidth]{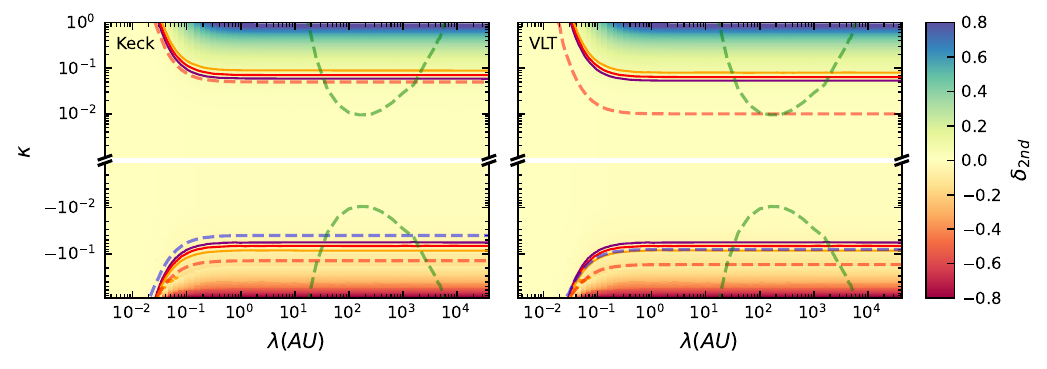}
\caption{
Constraints on $\kappa$ for varying $\lambda$ derived from future Sgr A* second photon ring deviation $\delta_{\rm 2nd}$ measurements. The red solid lines represent the upper and lower bounds for $\kappa$ when the resolution is improved by $50\%$ and measurements for second photon ring are expected to be $\delta_{\rm 2nd}=0.00^{+0.06}_{-0.07}$ ($\delta_{\rm 2nd} = 0.00\pm0.06$) for with Keck(VLTI) prior. Constraints from current \textit{EHT} Sgr A* measurements are shown by the blue and red dashed lines. The orange (purple) solid lines represent the constraints for $\kappa$ when the future resolution is improved by $20\%$ ($80\%$). For comparison, the constraints results from the S-stars measurements.\citep{2024PhRvD.109d4047T} are draw by the green dashed lines.
}
\label{constraint_SgrA_2nd}
\end{figure*}

\begin{figure}
\centering
\includegraphics[width=\textwidth]{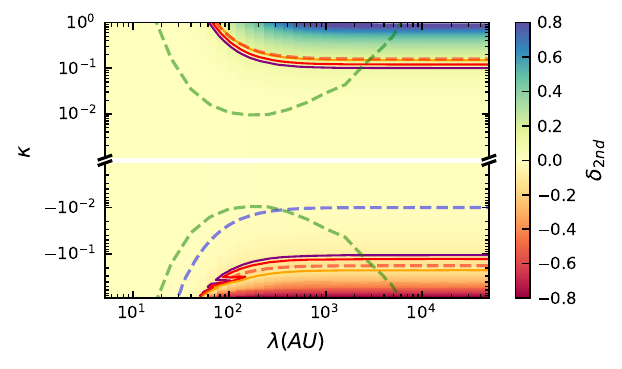}
\caption{
Legends are similar to that for Figure~\ref{constraint_SgrA_2nd}, except for M87* and $\delta_{\rm 2nd}=0.00\pm 0.11$ when the resolution is improved by $50\%$ (red lines).
}
\label{constraint_M87*_2nd}
\end{figure}

The next generation Event Horizon Telescope (\textit{ngEHT}) project aims to significantly advance the observational capabilities of the existing \textit{EHT} array \citep{2019BAAS...51g.256D}. The \textit{ngEHT} plans to add approximately $\sim$10 new observing sites worldwide by 2030, incorporating three simultaneous observing bands at $86$, $230$, and $345$\,GHz. These enhancements are expected to improve the angular resolution of current \textit{EHT} images by about $50$\%\citep{2023Galax..11...61J,2023arXiv231202130A}. 
{A space-baseline extension to the EHT, the Black Hole Explorer (BHEX) has been proposed recently. The BHEX is aiming at targeting the M87* and Sgr A* with radio-interferometric observations at frequencies of $100$\,GHz through $300$\,GHz and from an orbital distance of $\lesssim 30,000$\,km. This design will enable measurements of the first and second photon rings around both M87* and Sgr A* \citep{2020SciA....6.1310J}. Additionally, several studies indicate that the second rings of Sgr A* and M87* can be also distinctly characterized by using linear and circular polarization measurements \citep{2024ApJ...970L..24S, 2024arXiv241015325T}.}

We anticipate that detecting the second photon ring with this improved $50$\% angular resolution will provide an opportunity to further test theories of gravity.

For the second ring, we define a similar deviation parameter, i.e.,
\begin{equation}
\delta_{\rm 2nd} \equiv \frac{d_{\rm 2nd}}{d_{\rm Sch,2nd}}-1.
\end{equation}

For Sgr A*, under the assumption of a Schwarzschild metric, the future measurement for the second photon ring is expected to yield $\delta_{\rm 2nd} = 0.00^{+0.06}_{-0.07}$ with the Keck prior and $\delta_{\rm 2nd} = 0.00 \pm 0.06$ with the VLTI prior, as the resolution will be improved by approximately 50\% \citep{2018ASPC..517...15S, 2023Galax..11...61J, 2023arXiv231202130A}. The resulting constraints on the Yukawa parameters are illustrated in Figure~\ref{constraint_SgrA_2nd} (red solid lines). For scales $\lambda > 0.3$\,AU, both the Keck and VLTI priors yield constraints of $-0.07 < \kappa < 0.07$. As $\lambda$ decreases, the constraints on $\kappa$ become weak. Around $\lambda = 0.08$\,AU, which roughly corresponds to the Schwarzschild radius of Sgr A*, we find that $-0.12 < \kappa < 0.14$ for both priors. Even with further improvements in angular resolution, constraints within this range cannot be tightened.

For M87*, the future measurement for the second ring is expected to be $\delta_{\rm 2nd} = 0.00 \pm 0.11$. The resulting constraints on the Yukawa parameters are shown in Figure~\ref{constraint_M87*_2nd}. For $\lambda > 300$\,AU, we find $-0.14 < \kappa < 0.14$.

We also consider scenarios if the resolution of the next generation VLBI observation is improved by approximately $20\%$ ($80\%$) comparing with current \textit{EHT}, implying that the uncertainties of the measurement for $\delta_{\rm 2nd}$ should be reduced by factors of $1.2$ ($1.8$). The corresponding constraining results are shown by orange (purple) lines in Figure~\ref{constraint_SgrA_2nd}. For Sgr A*, we observe that both Keck and VLTI priors yield constraints of $-0.09 < \kappa < 0.09$ ($-0.06 < \kappa < 0.06$) when $\lambda > 0.3$\,AU for a $20\%$ ($80\%$) improvement. For M87*, we find $-0.25 < \kappa < 0.25$ ($-0.12 < \kappa < 0.12$) when $\lambda > 300$\,AU for a $20\%$ ($80\%$) improvement. Apparently, the constraining results are only slightly different from those obtained by assuming a $50\%$ improvement.

Unfortunately, there is no significant difference between the color maps derived from the second rings and those from the critical curves, suggesting that these two methods yield similar results. Nevertheless, the \textit{EHT} collaboration can directly detect only the peak intensity, rather than the critical curves. It is important to emphasize that the peak intensity is influenced not only by spacetime geometry but also by accretion physics. When determining the diameter of the critical curve, the \textit{EHT} collaboration adopted three types of simulations to assess the distribution of the calibration factor between these two factors \citep{2022ApJ...930L..17E}. This process carries substantial uncertainty because of our limited understanding of the accretion process. In contrast, the physical significance of the second ring is clearer: it is governed by unstable photon orbits and therefore is primarily dependent on the properties of the black hole spacetime \citep{2022ApJ...927....6B}. As such, we believe that this method could help mitigate some systematic uncertainties introduced by numerical simulations.

\section{Conclusions and Discussions}
\label{conclusions}

In this paper, we investigate the shadows of Yukawa black holes, focusing on how the Yukawa parameters $\lambda$ and $\kappa$ affect the shadow size. For a fixed $\lambda$, a decrease in $\kappa$ results in a smaller shadow, while an increase in $\kappa$ leads to a larger shadow. As $\lambda$ increases, the influence of $\kappa$ becomes more significant. When $\lambda$ reaches values on the order of hundreds of Schwarzschild radii, this influence stabilizes, maintaining a consistent level. These characteristics enable the constraint of the Yukawa parameters using \textit{EHT} observations. 

We then used the \textit{EHT} observations of M87* and Sgr A* to constrain $\lambda$ and $\kappa$. For Sgr A*, the Keck and VLTI observations provide different priors for the gravitational radius. For $\lambda>1$\,AU, the Keck prior Sgr A* shadow yields $\kappa=-0.04^{+0.09}_{-0.10}$, while the VLTI prior yields $\kappa=-0.08^{+0.09}_{-0.06}$. These constraints weaken as $\lambda$ decreases. At $\lambda=0.1$\,AU, the constraints become $-0.37<\kappa<0.17$ for the Keck prior and $-0.47<\kappa<0.04$ for the VLTI prior. For M87*, whose mass is significantly larger than the mass of Sgr A*, this system can only constrain $\kappa$ at higher values of $\lambda$. For $\lambda>1.5\times10^4$\,AU, the \textit{EHT} observation of M87* yields $\kappa=-0.01^{+0.17}_{-0.17}$. We find no deviation from the GR theory in this research so far. 

We then explore the potential constraints achievable by future observations, like \textit{ngEHT} and BHEX, of the second ring. Assuming a $\sim 50$\% improvement in resolution, the future measurement for the Sgr A* second ring is expected to yield $\delta_{\rm 2nd} = 0.00^{+0.06}_{-0.07}$ with the Keck prior and $\delta_{\rm 2nd} = 0.00 \pm 0.06$ with the VLTI prior. For scales $\lambda > 0.3$\,AU, both the Keck and VLTI priors yield constraints of $-0.07<\kappa<0.07$. As $\lambda$ decreases, the constraints on $\kappa$ weaken. At $\lambda = 0.08$\,AU, approximately the Schwarzschild radius of Sgr A*, the constraints become $-0.12<\kappa<0.14$ for both priors. For M87*, the future measurement of the second ring is expected to yield $\delta_{\rm 2nd} = 0.00 \pm 0.11$. For $\lambda > 300$\,AU, the constraints on $\kappa$ are $-0.14<\kappa<0.14$. If the future resolution of is improved by approximately $20\%$ (or $80\%$), the corresponding constraint results are presented in Figure~\ref{constraint_SgrA_2nd} [see orange (or purple) lines there]. For Sgr A*, both the Keck and VLTI priors provide constraints of $-0.09 < \kappa < 0.09$ ($-0.06 < \kappa < 0.06$) when $\lambda > 0.3$\,AU for a $20\%$ ($80\%$) improvement. For M87*, we find constraints of $-0.25 < \kappa < 0.25$ ($-0.12 < \kappa < 0.12$) when $\lambda > 300$\,AU for a $20\%$ ($80\%$) improvement.

{Our results show no significant difference between the color maps derived from the critical curve and the second photon ring, indicating that both methodologies yield comparable results. However, we argue that this conclusion relies on the assumption that the accretion physics is fully understood and that the GRMHD simulations used to interpret the observations are accurate. In practice, what the EHT measures directly are the locations of peak intensity in the image, not the critical curves themselves. The peak intensities in these shadow images are influenced not only by spacetime geometry but also by accretion physics. In determining the diameter of the black hole shadow via the critical curve, the \textit{EHT} collaboration utilized three distinct types of simulations to evaluate the distribution of the calibration factor between these two parameters \citep{2022ApJ...930L..17E}. Thus any unknown uncertainties or systematics in the simulations may bias the inferred radius of the critical curve. By contrast, the second photon ring arises from strong gravitational lensing and encodes information that depends primarily on the spacetime geometry, with only weak sensitivity to the uncertain astrophysical properties of the accretion flow \citep{2022ApJ...927....6B}. Moreover, it can be probed through direct measurements of linear and circular polarization, which are less affected by accretion model uncertainties \citep{2024ApJ...970L..24S, 2024arXiv241015325T}. Therefore, although both methods currently yield similar constraints, we include both in our analysis to highlight their conceptual distinction and to anticipate future observations—especially with instruments like the ngEHT and BHEX—that may resolve the second ring directly. 
}

Gralla (2021) \cite{2021PhRvD.103b4023G} pointed out that astrophysical uncertainties may significantly overshadow the effects induced by minor alterations in spacetime geometry. The EHT collaboration has considered multiple GRMHD simulations with different astrophysical concerns and metrics to derive $\delta$, the uncertainty in their measurements for Sgr A* and M87* should include both the astrophysical uncertainty and the metric uncertainty. In our work, we assume that this uncertainty dominates by the difference between the Yukawa-metric and the GR metric, and thus the limits on $\kappa$ and $\lambda$ we obtained may be conservative. More accurate constraints could be achieved by performing Yukawa-MHD simulation, but this is beyond the scope of this paper. 

{Various studies have estimated the Yukawa parameters using different approaches. For instance, the rotational curve serves as a key probe, as the original motivation for the Yukawa potential was to explain these curves \citep{Sanders1984}. Mota et al. (2011) \citep{Mota2011} used the rotational curve of low surface brightness galaxies and found $1.83 < \kappa < 11.67$ and $0.349 < \lambda < 75.810$\,kpc. Stabile \& Capozziello (2013) \citep{Stabile2013} obtained $-0.95 < \kappa < -0.92$ for $\lambda \sim 25-50$\,kpc. Moffat \& Rahvar (2013) \citep{Moffat2013} and Rahvar \& Mashhoon (2014) \citep{Rahvar2014} analyzed observed rotation curve datasets from several galaxies in The HI Nearby Galaxy Survey catalogue, finding $\kappa = -0.899 \pm 0.003$, $\lambda = 23.81 \pm 2.27$\,kpc and $\kappa = -0.916 \pm 0.041$, $\lambda = 16.95 \pm 8.04$\,kpc, respectively. In our study, Sgr A*'s shadow can only probe smaller scales due to its relatively short distance of $D \approx 8$\,kpc, yielding constraints of $-0.17 < \kappa < 0.05$. M87* lies much farther away at $D \sim 16.8$\,Mpc , and its shadow provides constraints of $-0.18 < \kappa < 0.16$ for $1.5\times10^4$\,AU$<\lambda<16.8$\,Mpc. It is important to note that this result should not be interpreted as a definitive rejection of the Yukawa potential, as the constraints on rotational curves are based on a specific set of parametrizations for the gas, bulge, disk, and dark matter profiles, and is not yet general enough to include all possibilities.

On the other hand, Tan \& Lu (2024) \citep{2024PhRvD.109d4047T} analyzed the orbits of S-stars in the Galactic Center, where the dynamics are much cleaner, and obtained constraints of $|\kappa| < 0.01$ for $\lambda \in (100,250)$\,AU and $|\kappa| < 2$ for $\lambda > 2 \times 10^4$\,AU. Additional constraints in other systems have been reported in \citep{Hees2017PRL, DeMartino2018PRD, DAddio2021PD, DeLaurentis2013MN, deLaurentis2015, Berti2011PRD, Choudhury2004APh}. Our results are derived from systems that probes smaller spatial scales and stronger gravitational potentials. The current \textit{EHT} constraints are even more stringent for $\lambda < 10$\,AU, and surpass those from S-star measurements by more than an order of magnitude for $\lambda > 1000$\,AU \citep{2024PhRvD.109d4047T}. Tan \& Lu (2024) \cite{2024PhRvD.109d4047T} argued that this weaken in constraints at large $\lambda$ in S‑star analyses arises from a cancellation between pericenter and apocenter precessions. In our case, all photons originate near the black hole and eventually reach the observer, thus this system probes the effects of the entire spacetime. These findings significantly advance the study of the Yukawa-type modifications to gravity and complement existing constraints from a range of astrophysical systems.

In our calculations, we adopt the Yukawa metric, derived in the framework of Verlinde's emergent gravity, which however perhaps pose an important conceptual challenge, i.e., BHs are not baryonic in the conventional sense. Verlinde interpreted BH as the point where all bits have been maximally coarse grained, thereby strongly influencing the surrounding entropy gradient \citep{2011JHEP...04..029V, 2017ScPP....2...16V}. In this view, observational constraints from SMBH shadows can be understood as constraints on the underlying entropy distribution. The inferred modifications influence the emergent geometry near compact objects regardless of the nature of the source, and can thus be probed observationally through shadow measurements.

Looking ahead, the next-generation \textit{ngEHT} will substantially improve the angular resolution of current \textit{EHT} images and expand the array’s accessible angular scales by an order of magnitude. In addition, the proposed BHEX mission will enable the detection of the second photon ring. Together, these advances hold great promise for refining constraints on the Yukawa parameters and testing alternative theories of gravity.
}

\section{Acknowledgments}
This work is supported by the National Natural Science Foundation of China (grant nos. 12273050 and 11991052), and the Strategic Priority Program of the Chinese Academy of Sciences (grant no. XDB0550300).

\end{document}